\documentclass[fleqn,10pt]{olplainarticle}
\usepackage{geometry}
\usepackage{times}
\usepackage{amsfonts}
\usepackage{xcolor}
\usepackage{setspace}
\definecolor{color2}{RGB}{255,255,255}

\title{Bayesian model for early dose-finding in phase I trials with multiple treatment courses}

\author[1,2]{Moreno Ursino} 
\author[3]{Lucie Biard\footnote{Corresponding author: {\sf{e-mail: lucie.biard@u-paris.fr}}}}
\author[3]{Sylvie Chevret }

\affil[1]{INSERM, Centre de Recherche des Cordeliers, Sorbonne Universit\'{e}, USPC, Universit\'{e} de Paris, 75006 Paris, France }
\affil[2]{F-CRIN PARTNERS Platform, AP-HP, Universit\'{e} de Paris, 75010 Paris, France.}
\affil[3]{Universit\'e de Paris, AP-HP, H\^opital Saint Louis, Service de Biostatistique et Information M\'edicale, INSERM U1153 Team ECSTRRA, Paris, France.}


\begin{abstract}
Dose-finding clinical trials in oncology aim to determine the maximum tolerated dose (MTD) of a new drug, generally defined by the proportion of patients with short-term dose-limiting toxicities (DLTs). Model-based approaches for such phase I oncology trials have been widely designed and are mostly restricted to the DLTs occurring during the first cycle of treatment, although patients continue to receive treatment for multiple cycles. 
We aim to estimate the probability of DLTs over sequences of treatment cycles 
via a Bayesian cumulative modeling approach, where the probability of DLT is modeled taking into account the cumulative effect of the administered drug and the DLT cycle of occurrence
. We propose a design, called DICE (Dose-fInding CumulativE), for dose escalation and de-escalation according to previously observed toxicities, which aims at finding the MTD sequence (MTS).
We performed an extensive simulation study comparing this approach to the time-to-event continual reassessment method (TITE-CRM) and to a benchmark. In general, our approach achieved a better or comparable percentage of correct MTS selection. Moreover, we investigated the DICE prediction ability.
\end{abstract}

\begin{document}

\flushbottom
\maketitle
\thispagestyle{empty}

\section{Introduction}

Dose-finding clinical trials in oncology of cytotoxic drugs, as well as of molecularly targeted agents, aim to determine the maximum tolerated dose (MTD) of a new drug, generally derived from the number of patients with short-term major treatment toxicities, e.g., dose-limiting toxicities (DLTs) observed in the trial. To better identify a safe dose with an acceptable toxicity profile, model-based approaches for such phase I oncology trials have been widely designed for the last 3 decades (see \cite{iasonos2014} for a review) and are mostly restricted to the DLTs occurring during the first or a few cycles of the studied drug. These designs usually ignore both the administration of the drug for multiple cycles and the time to occurrence of DLT. Moreover, the occurrence of DLT could be delayed, possibly related to the pharmacokinetics
 and/or to the cumulative doses of the drug. For example, \cite{postel14} showed in their research on phase I trials of molecularly targeted agents that DLTs occurred more frequently after the first cycle (1087 versus 936 at the first cycle). Moreover, they claimed that ``recommended phase 2 dose assessment should incorporate all available information from any cycle".
Thus, we were interested in approaches that extend dose-finding for phase I trials to handle the occurrence of DLTs over 
the different cycles of the treatment regimen.

The time-to-event continual reassessment method (TITE-CRM)~\citep{cheung2000}, which can be seen as a weighted version of the continual reassessment method~\citep{o1990}, was one of the first attempts to account for DLT in an extended follow-up window. 
In contrast, a few novel statistical models have been proposed to address the data from all the chemotherapy cycles in the estimation process, considering the effect of the dose yet to be eliminated from the body from the previous cycles, either via a simple pharmacokinetic model with fixed parameters \citep{legedza2000} or a cumulative logit formulation with a random intercept \citep{paoletti2015}. 
Other methods focus on the time to toxicity, and each administration is associated to a hazard contribution that is summed in estimating the cumulative effect of several cycles~\citep{braun05, braun07, liu09, zhang13}.
\cite{wages14} projected the dose schedule-finding task into a 2-dimensional problem and extended the partial order CRM
  developed for combination trials. \cite{lyu18} proposed a specific hybrid design, mixing an algorithm-based part with a Bayesian model-based side for sequences of doses over 3 cycles.

More recently, a Bayesian model for the conditional probability of experiencing a DLT at each cycle, given the lack of any DLTs on the previous cycles, has also been proposed~\citep{fernandes2016b}. The original model has three parameters, catching the effect of the first administration, the cumulative dose, and the body acquired resistance, respectively. However, a highly informative prior for the last parameter is necessary unless its estimate is biased, as shown in Appendix~A3.

We propose a cumulative modeling approach to phase I adaptive clinical trials, with the aims of (i) modeling 
the probability of DLT taking into account the cumulative effect of the administered drug and the DLT cycle of occurrence, and (ii) proposing a design for dose escalation and de-escalation according to estimated probabilities of DLT from this model.
In the next section, we describe the motivating trial. In Section 3, the new methodological approach is detailed. Simulation settings are shown in Section 4 followed by the results, which include the predictive ability of the model and the retrospective application to the motivating dataset. Last, comments are collected in the Discussion section.

\section{Data}
We used individual data
from a phase II clinical trial in myelodysplastic syndromes (MDS) to assess the DLT of lenalidomide over 5 cycles as an illustrative example. 
This trial (AZA-Rev, NCT00885508)  enrolled 49 patients, with 3 initial dose levels of lenalidomide of 10, 25 and 50 mg/day, in combination with azacitidine, 75 mg/m$^2$, with initial guesses of DLT of 0.2, 0.3 and 0.4, respectively. Cycles were scheduled to be administered every 4 weeks, at the same dose level in each patient, as long as there was no unacceptable toxicity or overt progression.
Although the trial was originally designed to identify the most successful safe dose according to the design proposed by \cite{zohar2006}, we retrospectively used toxicity data only to illustrate our setting, focusing on the 142 hematological adverse events that occurred along the 5 first cycles in 34 of these 49 patients.

\section{Methods}

Let $Y_{i,k}$ be the binary toxicity outcome (i.e., the realization $y_{i,k}$ is equal to 1 or 0 for DLT and no DLT, respectively) of the $i$th patient at the cycle $Z = k$, where $k \in {1, \ldots, K}$ and $K$ denotes the maximum number of cycles planned in the trial. Moreover, we define $\tilde{Y}_{i,k}$ as the marginal toxicity variable for the $i$th patient that takes the value 1 if the patient experiences toxicity 
between the inclusion and cycle $k$, and 0 otherwise. We also suppose that a patient can experience a maximum of one DLT since, after its occurrence, the patient drops out from the trial. Hereafter, we will use toxicity to refer to DLT.

Let $d_{i,k}$ be the dose given to the $i$th patient at cycle $k$, $\bar{s}_i$ be the theoretical sequence of doses given to the $i$th patient over the whole planned follow-up, provided the patient does not experience a DLT, and $D_{i,k}$ is the cumulative dose given to the patient excluding the loading dose (first dose), that is, $D_{i,k} = \sum_{m=2}^k d_{i,m}$ if $k >1$ and $D_{i,1} =0$ (at the first cycle). Let $\mathbf{\bar{s}} = \{\bar{s}_1, \ldots, \bar{s}_J  \}$ be the set of all the $J$ possible ordered sequences of doses that can be used in the panel, where $\bar{s}_j = \{s_{1,j}, \ldots, s_{K,j}  \}$, and $s_{k,j} \leq s_{k,j+1}$ for all $k \in {1, \ldots, K}$ and $j \in {1, \ldots, J-1}$ and $\bar{s}_j \neq \bar{s}_{j+1}$.

We propose to directly model the cumulative probability of toxicity at cycle $k$, that is, $P(\tilde{Y}_{i,k}=1) = \sum_{m=1}^k P(Y_{i,m}=1)$, with  $P(Y_{i,m}=1) = P(Y_{i,m}=1 | Y_{i,m-1}=0) P(Y_{i,m-1}=0 | Y_{i,m-2}=0) \cdots P(Y_{i,1}=0)$ . Let 
\begin{equation}
 P(\tilde{Y}_{i,k} =1) = 
f_{\theta}(d_{i,1}, k, D_{i,k}),
\end{equation}
with $f_{\theta}(.)$ denoting an appropriate parametric link function. Following this notation, the probability that a patient experiences a DLT exactly at cycle $k$ is given by
\begin{equation}\label{eq:tox}
 P(Y_{i,k}=1) =  P(\tilde{Y}_{i,k} =1) -  P(\tilde{Y}_{i,k-1} =1),
\end{equation}
with the convention that $P(\tilde{Y}_{i,0} =1)=0$, while the probability that a patient does not experience toxicity at cycle $k$ can be computed as 
\begin{equation}\label{eq:notox}
 P(Y_{i,k}=0) = 1- P(\tilde{Y}_{i,k} =1).
\end{equation}
Proofs of \eqref{eq:tox} and \eqref{eq:notox} are given in Appendix~A1.
Therefore, at each stage of follow-up, the $i$th patient will contribute to the likelihood up to the maximum completed treatment cycle, $k_i \in \{1, \ldots, K\}$, as
\begin{align}
\mathcal{L}_i 
(\theta| \mathbf{y} ) =& 
 P(Y_{i,k_i}=0)^{1-y_{i,k_i}} P(Y_{i,k_i}=1)^{y_{i,k_i}} \nonumber \\
 =& \left(1- P(\tilde{Y}_{i,k_i} =1)\right)^{1-y_{i,k_i}} \left(P(\tilde{Y}_{i,k_i} =1) -  P(\tilde{Y}_{i,k_i-1} =1)\right)^{y_{i,k_i}} \nonumber \\
 =&  \left(1- f_{\theta}(d_{i,1}, k_i, D_{i,k_i-1})\right)^{1-y_{i,k_i}}  \left(f_{\theta}(d_{i,1}, k_i, D_{i,k_i})-  f_{\theta}(d_{i,1}, k_{i-1}, D_{i,k_i-1}) \right)^{y_{i,k_i}}. 
\end{align}
The likelihood for $n$ accrued patients is, as usual, $\mathcal{L} 
(\theta| \mathbf{y} ) = \prod_{i=1}^n \mathcal{L}_i 
(\theta|\mathbf{y} )$.

Regarding the definition of $f_{\theta}(.)$, we propose an extension of the well-known Bayesian logistic regression model~\citep{Neuenschwander08} adapted to model a cumulative dose effect. Specifically, 
\begin{equation}
 \mbox{logit}\left( P(\tilde{Y}_{i,k} =1)\right) = \alpha + \exp (\beta) \log \left(\frac{d_{i,1}}{d^*}\right) + \exp(\gamma) \log\left(\frac{D_{i,k}}{D^*}+1\right) g(k),
\end{equation}
where $d^*$ denotes the reference dose, which usually is the best guess at the MTD, $D^*$ refers to the cumulative reference dose, that is, the final cumulative dose (at the last cycle) in the reference sequence $\bar{s}^*$, and the $+1$ term allows a nonnegative covariate effect along with $\exp (\gamma)$. In this setting, $\theta = \{\alpha, \beta,\gamma\}$.
The cycle is used as multiplicative factor $g(k)$ to preserve the monotonicity of the toxicity versus the sequences in the panel and to be able to distinguish two sequences with the same cumulative doses but different single doses at each cycle. The three parameters of the model have a specific influence:  $\alpha$ is linked to the probability of toxicity at the reference dose $d^*$ at cycle 1, $\exp(\beta)$ controls the shape of the toxicity sigmoid curve at cycle 1 and $\exp(\gamma)$ catches the nonnegative increment in toxicity risk due to the cumulative doses over cycles.

Independent normal prior distributions can be associated with the three parameters, that is, $\alpha \sim \mathcal{N}(\mu_\alpha, \sigma_\alpha)$, $\beta \sim \mathcal{N}(\mu_\beta, \sigma_\beta)$  and $\gamma \sim \mathcal{N}(\mu_\gamma, \sigma_\gamma)$. Under Bayesian inference, the posterior distribution of the parameters is obtained as $p_{post}(\theta) \propto  \mathcal{L} (\theta | \mathbf{y} ) p_{prior}(\theta)$.

\subsection{Dose-finding design}

The probability of toxicity associated with the theoretical dose-sequence $j$ of the panel is computed as
\begin{equation}
p_{T}(\bar{s}_j) = f_{\theta}\left(s_{j,1}, K, \sum_{m=2}^{K} s_{j,m} \right).
\end{equation}
Upon accrual of the first patient of the next cohort, the model parameters are estimated using all available patient data, including incomplete observations from those patients who are still in the follow-up window (that is, at cycle $k_i < K$ without having experienced a DLT), and  $\hat{p}_{T}$ is estimated using the posterior distribution of the parameters, $p_{post}(\alpha, \beta, \gamma)$.
The dose-sequence allocated to the next cohort of patients will be that with estimated probability of toxicity closest to the target $\pi$, that is, 
\begin{equation}\label{eq:nextdose}
\bar{s}_{next} = \mbox{arg} \min_{j} | \hat{p}_{T}(\bar{s}_j) - \pi|.
\end{equation}
Several estimators of $\hat{p}_{T}$ can be considered; 
 we suggest using 
the predictive probability of toxicity, that is, the expected value of $p_{T}(\bar{s}_j) $ with respect to the posterior distribution ($E_{p_{post}(\alpha, \beta, \gamma)}[p_{T}(\bar{s}_j)]$) or the median of the predicted probabilities, that is, the median of the  posterior distribution of $p_{T}(\bar{s}_j) $ computed from  the parameter posterior distribution.

The estimated MTD sequence (MTS) at cycle $k$ is defined by that sequence with estimated probability of toxicity closest to the target $\pi$, that is, 
\begin{equation}\label{eq:MTS}
    \mbox{MTS}_k= \mbox{arg} \min_{j} | \hat{p}_{T}(\bar{s}_{k,j}) - \pi|,
\end{equation}
where $\bar{s}_{k,j}$ denotes the sequence $\bar{s}_{j}$ truncated at cycle $k$.
At the end of the trial, the estimated MTS is thus MTS$_K$. 

The first cohort of patients is allocated to the first sequence in the dose panel, $\bar{s}_1$.
No-skipping rules apply, that is, a new dose sequence $j$ cannot be assigned to a cohort of patients if dose sequence $j-1$ has not already been given to any previous patient. 
Regarding safety constraints, we impose a safety stopping rule on the posterior probability of toxicity of the first dose sequence. At any stage of the dose-finding study, the trial is stopped if:
\begin{equation} \label{eq:stopping_rule}
P_{post}\left(p_{T}(\bar{s})_1> \pi \right) > \tau_T,
\end{equation}
where  $\tau_T$ refers to a prespecified threshold, usually between 0.8 and 0.99.
This proposed method will be referred to as DICE (Dose-fInding CumulativE) hereafter.

\section{Simulations}

\begin{figure}
    \centering
    \includegraphics[width=\textwidth]{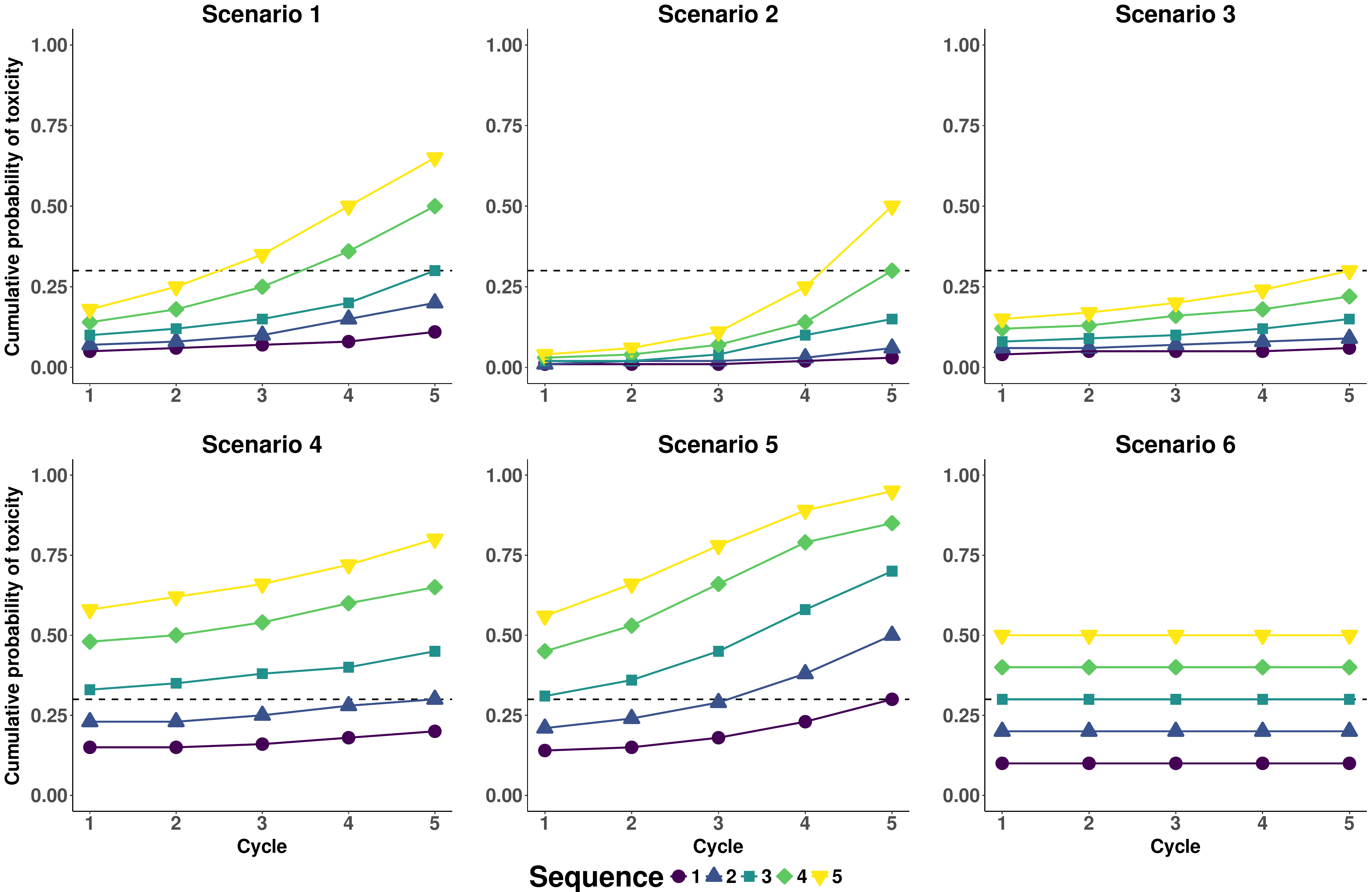}
    \caption{Cumulative probabilities of toxicity for each scenario and sequence.}
    \label{fig:scen_plot}
\end{figure}

We evaluated the performance of the proposed method in six scenarios that differed in terms of the five dose sequences over the five cycles. Dose sequences were defined by repeating the same dose level $d \in$ \{5, 7, 10, 15, 20 mg\}, at each cycle; in other words, there was no intrapatient dose escalation/de-escalation across cycles.  Figure~\ref{fig:scen_plot} displays the cumulative probabilities of toxicity at each cycle for each dose sequence, with exact probabilities of toxicity detailed in Appendix~A2, Table~\ref{tablescen}. 
Scenarios resulted in MTD sequences at different panel positions according to the target $\pi=0.3$. Briefly, in Scenario 1, the cumulative probabilities over the 5 cycles indicate that for the dose sequence 3, which is where the trial begins with administration of dose 3, the cumulative probability at the end of the cycles is the closest to the target of 0.3; this dose sequence is thus the MTD sequence for that scenario. In the other scenarios, the MTD sequence differs from Scenario 5 (with dose sequence 1), to Scenario 3 (where the MTS is the fifth).
In particular, in Scenario 6, we examined the situation where DLT can be seen only at the first cycle. 

We simulated individual toxicity outcomes via the conditional probabilities of toxicity corresponding to the cumulative probabilities that defined the scenarios; we generated outcomes for all dose sequences and all cycles for each trial. Each simulated dataset was stored, and patient outcomes were obtained from this complete dataset. Using this approach, when two compared methods coincide in dose allocation, the same patient outcomes are used in both estimation runs, rendering comparison across methods straightforward. 

The threshold $\tau_T$ was set equal to 0.9 for the safety stopping rule (equation \ref{eq:stopping_rule}) that applied when at least six patients had already been accrued. We used a cohort size of 1 or 3 subjects and a fixed sample size of 30.

\begin{figure}
    \centering
    \includegraphics[width=\textwidth]{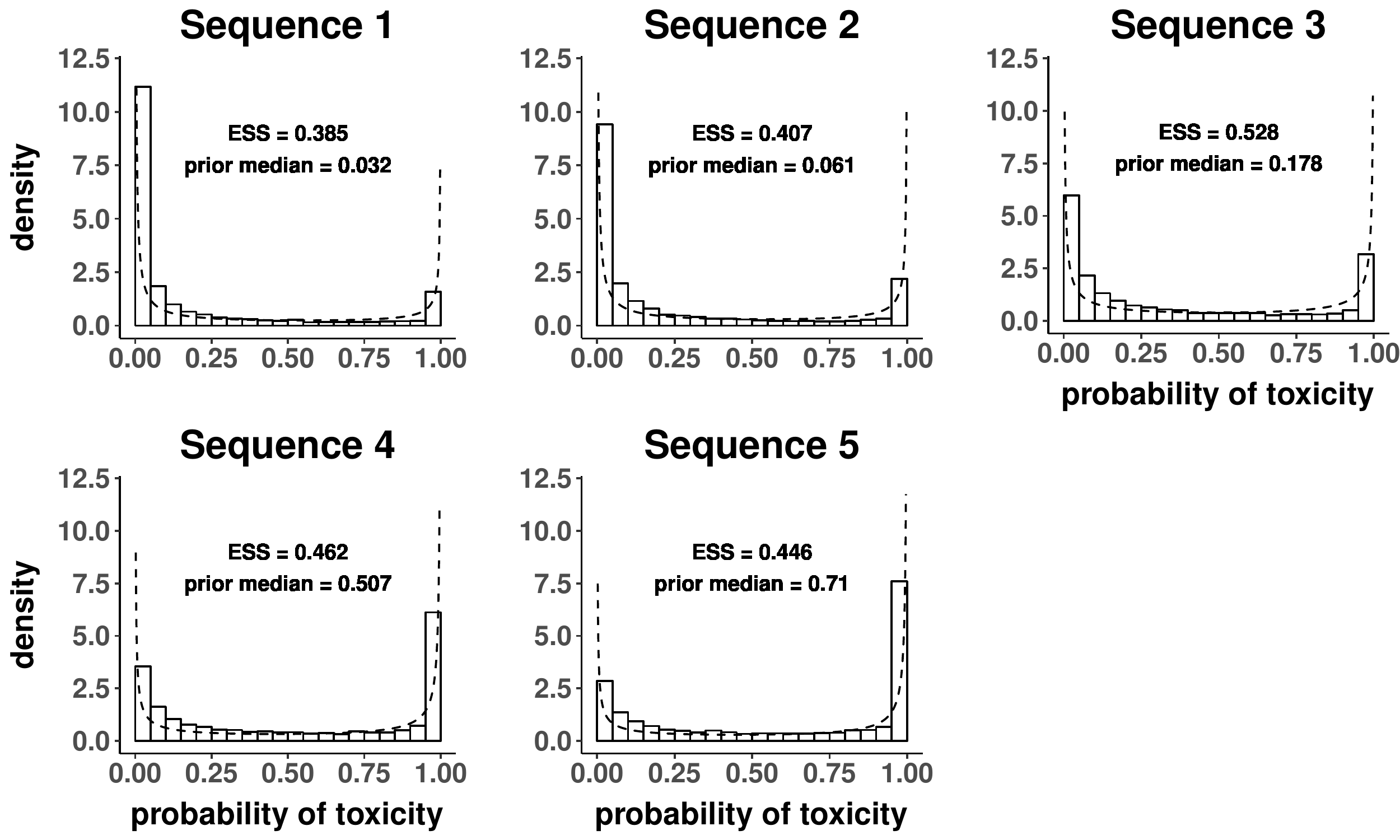}
    \caption{Induced prior probability of toxicity at the last cycle for each sequence along with the approximated effective sample size (ESS) and the median prior value. The dashed line represents the fitted beta distribution.}
    \label{fig:scen_prior}
\end{figure}

The function $g(k)$ was defined as $g(k) = k/K$ and, after a sensitivity analysis (not shown here), the following prior distributions were selected: $\alpha \sim \mathcal{N}(-3, 2)$, $\beta \sim \mathcal{N}(0, 2)$ and $\gamma \sim \mathcal{N}(0, 2)$. For computational stability, we truncated the normal distribution associated with $\alpha$ in the interval [-10, 5]. The induced probability of toxicity at cycle 5 for each schedule, along with the approximated effective sample size (ESS) and the prior median, are shown in Figure~\ref{fig:scen_prior}. The ESS was computed using the crude method proposed by \cite{Morita08}. The simulated prior probability of toxicity at each schedule $j$ was approximated by a beta distribution, $Beta(a_j, b_j)$, and the ESS was computed as $a_j + b_j$.
The median of the induced posterior distribution of $p_{T}(\bar{s}_j) $ by the parameters posterior distribution was used to estimate $\hat{p}_{T}$.

We compared the DICE proposal to the TITE-CRM~\citep{cheung2000} and to a benchmark. 
First, a logistic TITE-CRM was implemented using the \texttt{dfcrm} package in R software. For this design, we considered an observation window corresponding to the five dosing cycles, with the TITE-CRM MTD being equivalent to the DICE MTS$_K$. To make comparable designs, we added stopping rules to the TITE-CRM using the estimated confidence interval of toxicity probabilities given by the function. In detail, the trial is stopped if the lower bound of the $100\times(1-2\times(1-\tau_T))$~\% confidence interval of the first dose is higher that $\pi$ or if it is not computable. The skeleton was chosen using the method of indifference intervals with the prior MTD at the third place of the dose sequence panel and a half width of the indifference intervals of 0.10 \citep{lee2009}.

Then, a benchmark was defined using the estimated empirical probabilities using all simulated data at all doses for all patients in each trial. In each simulated trial, the probability of toxicity at each sequence of the panel was estimated as 
\begin{equation}\label{eq:benchmark}
\tilde{p}_{tox}(\bar{s}_j) = \frac{n_{DLT, j}}{n},
\end{equation}
where $n_{DLT,j}$ refers to the number of patients who experienced toxicity at dose sequence $j$ and $n$ the trial sample size. In agreement with Eq.~\ref{eq:nextdose}, we defined the MTD as $\mbox{arg} \min_{j} | \tilde{p}_{T}(\bar{s}_j) - \pi|.$

For each scenario, 5,000 independent trials were simulated.
 As measures of method performance, we computed, for each scenario, the proportion of the correct MTD sequence selection (PCS), the proportion of dose sequence allocation, and the median [interquartile range] of the observed toxicities over the trial. 

\section{Results}
\subsection{Operating characteristics}
Table~\ref{tableres1} shows the simulation results in terms of the proportion of selection by 
the MTD sequence, the proportion of dose allocation and the  number of DLTs per trial for the proposed method, the TITE-CRM and the benchmark. For the DICE and the TITE-CRM, the results are given for both cohorts of 1 and of 3 patients. 

In Scenario 1, the percentage of the correct MTS selection (PCS) for the DICE was 65\% and 62\% for cohorts of size 1 and 3, respectively, compared to 57\% and 53\% for the TITE-CRM, respectively, and 58\% for the benchmark.
 The TITE-CRM with a cohort of size 3 has the lowest number of median DLTs but also the lowest percentage of patients allocated to the MTD, 33\%. 

In Scenario 2, the PCS of DICE ranged from 65\% to 70\%, close to those of TITE-CRM that ranged from 64\% to 68\%, while the benchmark selected the MTS 76\% of the time. 

In Scenario 3, the MTS was located at the end of the dose panel; the PCS of the DICE was 63\% and 66\% for cohorts of size 1 and 3, respectively, while those of TITE-CRM were 33\% and 50\%, respectively, and that of the benchmark, 49\%. Median numbers of DLTs were comparable, ranging from 5 to 7. 

Scenario 4 showed PCS ranging from 44\% to 56\%, with the lowest achieved by the DICE with a cohort of 1. 

In Scenario 5, the PCS of DICE and TITE-CRM were both close to 70\%
, with approximately 10\% of trials stopped early for safety reasons or ended without selection of a schedule as MTS. The benchmark, which does not incorporate stopping rules, had a PCS of 92\%.

In the last Scenario 6, the PCS ranged between 44\% and 54\%, while the median number of DLTs was between 7 and 9.

\begin{center}
\begin{table}
\caption{Results of the 5000 simulated trials for the six proposed scenarios for each method. The proportion of the selected MTD sequence, proportion of dose allocation and number of DLTs are shown. In bold, the results regarding the true MTD sequence in each scenario are shown. }\label{tableres1}
\small{\begin{tabular}{lllllllllllll}
\hline
\textbf{Method}   & \multicolumn{6}{c}{\textbf{Dose sequence selection}} & \multicolumn{5}{c}{\textbf{Dose sequence allocation}} & \textbf{DLTs} \\ 
\textbf{- cohort} & None & 1 & 2 & 3 & 4 & 5 & 1 & 2 & 3 & 4 & 5 & median (Q1, Q3) \\ 
\hline
  Scenario 1  & &  &  & &  & &  &  &  &  &  &  \\  
  $p_{tox}$ & & 0.11 & 0.20 & \textbf{0.30} & 0.50 & 0.65 &  &  &  &  &  \\  
  DICE - 1 & 0.003 & 0.012 & 0.199 & \textbf{0.65} & 0.133 & 0.002 & 0.078 & 0.19 & 0.441 & 0.192 & 0.099 & 10 (9, 12) \\ 
  TITE-CRM - 1 & 0.004 & 0.023 & 0.3 & \textbf{0.571} & 0.1 & 0.002 & 0.143 & 0.271 & 0.363 & 0.153 & 0.069 & 9 (7, 11) \\ 
  DICE - 3 & 0.001 & 0.014 & 0.194 & \textbf{0.623} & 0.157 & 0.01 & 0.145 & 0.245 & 0.395 & 0.166 & 0.049 & 9 (8, 10) \\ 
  TITE-CRM - 3 & 0.003 & 0.026 & 0.352 & \textbf{0.533} & 0.085 & 0.001 & 0.211 & 0.333 & 0.334 & 0.106 & 0.016 & 7 (6, 9) \\ 
  benchmark & 0 & 0.043 & 0.336 & \textbf{0.578} & 0.043 & 0.001 &  &  &  &  &  & \vspace{2mm} \\ 
  Scenario 2  & &  &  & &  & &  &  &  &  &  &  \\  
  $p_{tox}$ & & 0.03 & 0.06 & 0.15 & \textbf{0.30} & 0.50 &  &  &  &  &  \\  
  DICE - 1 & 0 & 0 & 0.002 & 0.186 & \textbf{0.702} & 0.109 & 0.037 & 0.048 & 0.2 & 0.41 & 0.304 & 9 (8, 11) \\ 
  TITE-CRM - 1 & 0 & 0 & 0.004 & 0.208 & \textbf{0.68} & 0.108 & 0.046 & 0.066 & 0.211 & 0.397 & 0.28 & 9 (8, 10) \\ 
  DICE - 3 & 0 & 0 & 0.004 & 0.167 & \textbf{0.659} & 0.169 & 0.106 & 0.122 & 0.217 & 0.326 & 0.229 & 8 (7, 9) \\ 
  TITE-CRM - 3 & 0 & 0 & 0.004 & 0.236 & \textbf{0.644} & 0.116 & 0.113 & 0.13 & 0.255 & 0.341 & 0.161 & 7 (6, 8) \\ 
  benchmark  & 0 & 0 & 0.008 & 0.183 & \textbf{0.756} & 0.053 &  &  &  &  &  &  \vspace{2mm} \\ 
  Scenario 3  & &  &  & &  & &  &  &  &  &  &  \\  
  $p_{tox}$ & & 0.06& 0.09 & 0.15 & 0.22 & \textbf{0.30} &  &  &  &  &  \\  
  DICE - 1 & 0 & 0 & 0.007 & 0.15 & 0.216 & \textbf{0.626} & 0.045 & 0.056 & 0.183 & 0.194 & 0.521 & 7 (6, 8) \\ 
  TITE-CRM - 1 & 0.001 & 0.001 & 0.025 & 0.145 & 0.33 & \textbf{0.499} & 0.075 & 0.098 & 0.176 & 0.25 & 0.401 & 6 (5, 8) \\
  DICE - 3 & 0 & 0 & 0.004 & 0.089 & 0.242 & \textbf{0.664} & 0.11 & 0.12 & 0.206 & 0.216 & 0.348 & 6 (5, 7) \\ 
  TITE-CRM - 3 & 0.001 & 0.001 & 0.046 & 0.229 & 0.393 & \textbf{0.33} & 0.147 & 0.184 & 0.261 & 0.238 & 0.17 & 5 (4, 6) \\ 
  benchmark  & 0 & 0.004 & 0.017 & 0.115 & 0.373 & \textbf{0.492} &  &  &  &  &  & \vspace{2mm}  \\ 
  Scenario 4  & &  &  & &  & &  &  &  &  &  &  \\  
  $p_{tox}$ & & 0.20 & \textbf{0.30} & 0.45 & 0.65 & 0.80 &  &  &  &  &  \\  
  DICE - 1 & 0.015 & 0.197 & \textbf{0.444} & 0.333 & 0.011 & 0.001 & 0.234 & 0.334 & 0.353 & 0.058 & 0.021 & 11 (9, 12) \\ 
  TITE-CRM - 1 & 0.038 & 0.26 & \textbf{0.54} & 0.16 & 0.003 & 0 & 0.411 & 0.374 & 0.169 & 0.033 & 0.012 & 9 (7, 10) \\ 
  DICE - 3 & 0.01 & 0.199 & \textbf{0.477} & 0.299 & 0.015 & 0.001 & 0.274 & 0.375 & 0.291 & 0.051 & 0.009 & 10 (9, 11) \\ 
  TITE-CRM - 3 & 0.029 & 0.284 & \textbf{0.557} & 0.128 & 0.002 & 0 & 0.483 & 0.383 & 0.12 & 0.013 & 0.001 & 8 (7, 9) \\ 
  benchmark  & 0 & 0.339 & \textbf{0.558} & 0.103 & 0 & 0 &  &  &  &  &  &  \vspace{2mm} \\ 
  Scenario 5  & &  &  & &  & &  &  &  &  &  &  \\  
  $p_{tox}$ & & \textbf{0.30 }& 0.50 & 0.70 & 0.85 & 0.95 &  &  &  &  &  \\  
  DICE - 1 & 0.144 & \textbf{0.688} & 0.163 & 0.005 & 0 & 0 & 0.585 & 0.272 & 0.106 & 0.025 & 0.012 & 12 (10, 13) \\ 
  TITE-CRM - 1 & 0.158 & \textbf{0.721} & 0.121 & 0 & 0 & 0 & 0.723 & 0.201 & 0.048 & 0.018 & 0.009 & 10 (9, 12) \\ 
  DICE - 3 & 0.11 & \textbf{0.72} & 0.166 & 0.004 & 0 & 0 & 0.611 & 0.299 & 0.082 & 0.008 & 0 & 11 (10, 13) \\ 
  TITE-CRM - 3 & 0.131 & \textbf{0.737} & 0.131 & 0.001 & 0 & 0 & 0.743 & 0.222 & 0.033 & 0.002 & 0 & 10 (8, 12) \\ 
  benchmark  & 0 & \textbf{0.929} & 0.07 & 0 & 0 & 0 &  &  &  &  &  &  \vspace{2mm} \\ 
  Scenario 6  & &  &  & &  & &  &  &  &  &  &  \\  
  $p_{tox}$ & & 0.10 & 0.20 & \textbf{0.30} & 0.40 & 0.50 &  &  &  &  &  \\  
  DICE - 1 & 0 & 0.012 & 0.156 & \textbf{0.534} & 0.239 & 0.059 & 0.084 & 0.179 & 0.436 & 0.198 & 0.102 & 9 (8, 10) \\ 
  TITE-CRM - 1 & 0.005 & 0.022 & 0.337 & \textbf{0.445} & 0.171 & 0.021 & 0.188 & 0.321 & 0.307 & 0.14 & 0.045 & 7 (6, 9) \\ 
  DICE - 3 & 0 & 0.013 & 0.151 & \textbf{0.507} & 0.248 & 0.081 & 0.134 & 0.224 & 0.401 & 0.165 & 0.075 & 8 (7, 10) \\ 
  TITE-CRM - 3 & 0.003 & 0.022 & 0.403 & \textbf{0.434} & 0.127 & 0.011 & 0.248 & 0.384 & 0.275 & 0.081 & 0.012 & 7 (5, 8) \\ 
  benchmark  & 0 & 0.015 & 0.294 & \textbf{0.489} & 0.172 & 0.031 &  &  &  &  &  &  \\  
\hline
\end{tabular}}
\end{table}
\end{center}

\subsection{Predictions}
\begin{figure}
    \centering
    \includegraphics[width=\textwidth]{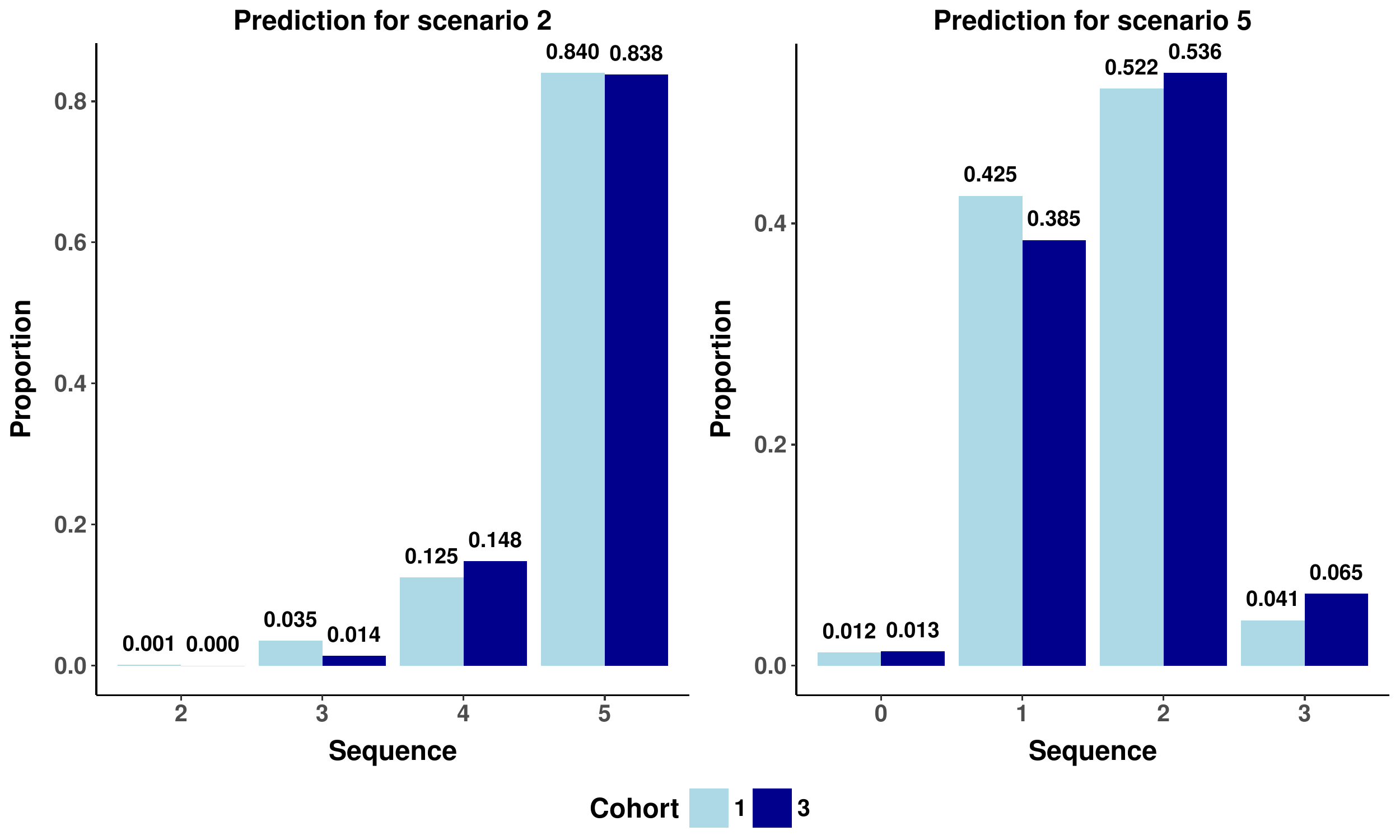}
    \caption{Simulation study: Probability of sequence selection using the model predictions at earlier cycles than the predefined window. Left panel: selection based on predicted cumulative probability of toxicity at cycle 4 in scenario 2; Right panel: selection based on predicted cumulative probability of toxicity at cycle 3 in scenario (x-axis 0 indicating no sequence recommendation at the end of the trial)
    }
    \label{pred}
\end{figure}

Using the DICE estimation at the end of the trial, clinicians can make predictions of toxicity probability associated with sequences with a lower number of cycles, that is, MTS$_k$ with $k<K$. Figure~\ref{pred} shows the proportions of MTD sequence selection in the case of only 4 cycles in Scenario 2, that is, MTS$_4$, and  MTS$_3$ for Scenario 5, with only 3 cycles. In this shortened Scenario 2, the true MTS$_4$ is 
schedule 5 and the DICE has a PCS of 84\%, regardless of the cohort size, while it is approximately 53\% for the shortened Scenario 5, where the true MTS$_3$ is schedule 2.

\subsection{Application to the motivating example}

We applied the DICE model retrospectively to the AZA-Rev dataset. We set the same prior distribution and the same $g(k)$ chosen for the simulation study, that is, $g(k) = k/K$ and $\alpha \sim \mathcal{TN}_{[-10,5]}(-3, 2)$, $\beta \sim \mathcal{N}(0, 2)$ and $\gamma \sim \mathcal{N}(0, 2)$, where $\mathcal{TN}$ denotes the truncated normal distribution. 
As the reference dose sequence,
we chose the one in the middle of the dose panel, that is, 25 mg/day given at each cycle.

\begin{figure}
    \centering
    \includegraphics[width=\textwidth]{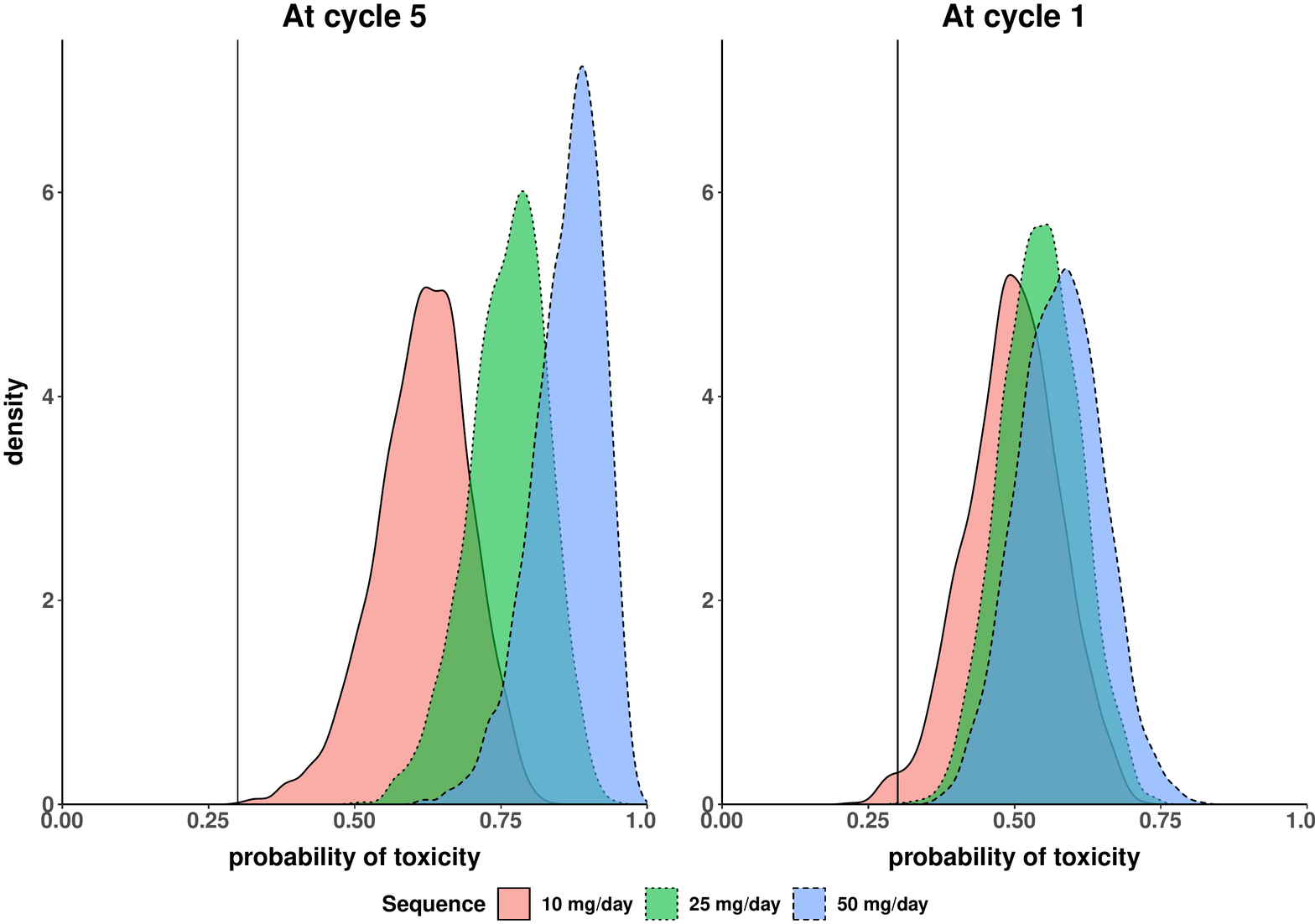}
    \caption{Results of retrospective analysis of AZA-Rev trial. Posterior probabilities of toxicity are plotted for each sequence at cycle 5, left panel, and at cycle 1, right panel.}
    \label{res:aza}
\end{figure}

Figure~\ref{res:aza} shows the posterior probability of toxicity for each dose schedule at the last and at the first cycle. The median of the posterior probability of toxicity is equal to 62.1\%, 76.8\% and 87.2\% for 10, 25 and 50 mg/day, respectively, at cycle 5. Therefore, $\mbox{MTS}_5$ is the first dose sequence. However, applying the stopping rule of Eq.~\ref{eq:stopping_rule}, the probability that the first dose sequence is too toxic is 99.8\%. Thus, $\mbox{MTS}_5$ should not be selected as a safe sequence.
At cycle one, the probabilities are closer together, with 49.9\%, 54.3\%, and 57.9\%. Again, the stopping rule imposes no $\mbox{MTS}_1$ selection since $P_{post}(p_{T}(\bar{s})_1> 0.3) = 97.0\%$.

\section{Discussion}

We aimed at extending dose-finding methods for phase I trials that could handle the occurrence of DLTs over the different cycles of a treatment regimen. Therefore, we proposed a cumulative modeling approach, the DICE, with a new quantity of interest, the MTD sequence (MTS). Indeed, accounting for repeated dosing of the treatment over multiple cycles, the MTS extended the concept of MTD by defining not a single but a dose sequence with a cumulative probability of DLT closest to the target (e.g. 25\%); thus, the MTS is strictly linked to the number of cycles considered for dose finding, since considering a different number of cycles (reduced or increased) could lead to a different sequence choice in the candidate dose panel. The DICE showed comparable operational characteristics to the TITE-CRM, with a higher proportion of correct dose selection when the MTS was located in the second half of the schedule panel, that further allowed computing predictions over the treatment cycles evaluated by the trial. 

Predictions, that is, computation of probabilities of toxicity at different sequences and cycles, is indeed a key point of the new method. For instance, should the treatment reveal efficacy  with a lower number of cycles in later assessments, the DICE would be able to estimate the new probabilities of toxicity and the corresponding new shortened MTS. Although DICE could theoretically also provide predictions for untested sequences, such as different doses or a larger number of cycles, relying on the proposed parametric working model, we should consider with great caution such predictions involving doses or cycles never used in the estimation process (a sort of extrapolation).

Most other methods proposing toxicity risk predictions have stronger underlying assumptions than DICE. Moreover, cycle effect is usually not taken into account~\citep{legedza2000, braun05, braun07, liu09, zhang13} or a random effects structure is imposed (resulting in a more complex model with the associated computational issues). On the other hand, the \cite{fernandes2016b} method, working on conditional probabilities, can capture both the effect of resistance to the drug acquired by individuals upon repeated intakes and of the cumulative dose. However, as shown in Appendix~A3, a highly informative prior for the last parameter in their working model is necessary; otherwise, its estimate is biased and unreliable probability estimations arise.

A limitation regarding the dose-allocation rule is the need of ordered sequences, in terms of toxicity risk. If we cannot state a clear order, the nonskipping rule should be rethought. On the other hand, the DICE is based on the underlying assumption of the higher the cumulative dose, the higher the toxicity probability. Even if the model allows to theoretically capture the absence of effect of the cumulative dose, with $\gamma = 0$, it should not be applied if an increasing toxicity followed by a plateau is expected.  In our setting, we supposed that the same dose is administered at each cycle, which is a common and simple therapeutic regimen. However, the model remains valid with intrapatient dose adjustment, when different doses may be administered at different cycles. It can also accommodate dose mistakes given to the patient, given the actually instead of the scheduled administered doses can be used in the model. Nevertheless, the model assumes that the probability of toxicity at the final cycle $K$ corresponds to a given cumulative dose, regardless of the regimen followed to achieve it.

In our simulation study, whenever the stopping rule was fulfilled, the trial accrual was considered as stopped. However, in a real trial, we suggest only suspending accrual and waiting to obtain complete follow-up of the last patients and then re-estimate the model. If the stopping rule still holds, the trial should end; otherwise, accrual can start again. 

Of note, in some scenarios, the DICE outperformed the benchmark in terms of correct sequence selection. This has been observed previously for the same benchmark approach with a binary endpoint and the CRM \citep{Cheung_book}. In the same manner, we hypothesize parametric working models might be able to inform the dose-toxicity relationship and outperform a nonparametric method in dose selection, given the limited sample sizes of phase I trials. These considerations all the more apply to the present setting where the working model relies on richer information (cycles, cumulative dose) compared to the benchmark for basic binary endpoints (Eq. \ref{eq:benchmark}). Consistently, the benchmark improved in simulations with larger sample sizes (data not shown).

In conclusion, we developed DICE, a dose-finding clinical trial design for treatments with multiple cycles. The DICE targets the cumulative probability of toxicity over the course of treatment and incorporates the potential effect of cumulative doses. It showed desirable operating characteristics compared to the TITE-CRM. Moreover, the DICE can be applied to account for intrapatient dose adjustments.

\section*{Acknowledgments}

\subsection*{Financial disclosure}

None.

\subsection*{Conflict of interest}

The authors declare no potential conflict of interests.



\section*{Appendix}

\subsection*{A.1.\enspace Derivation of likelihood contributions}\label{app1}
Derivation of likelihood contributions, that is, Eq.~\ref{eq:tox} and \ref{eq:notox}, for the $i$th patient. For the sake of clarity, let $p_k$ denote the conditional probability of toxicity at cycle $k$ given not having observed any DLT in previous cycles.  By definition,
\begin{align*}
  P(\tilde{Y}_{i,k}=1) &= \sum_{q=1}^{k}  P(Y_{i,q}=1) \\
 &= \begin{cases} 
 p_1, & \mbox{if } k =1  \\ 
 p_1 + \sum_{q=2}^{k} p_q \prod_{m=1}^{q-1} (1 - p_m), & \mbox{if } k > 1 
 \end{cases}  
\end{align*}
Therefore, the probability of toxicity at given specific cycles, $k \in {1, \ldots, K}$ is obtained as $   P(Y_{i,1}=1) = P(\tilde{Y}_{i,1}=1) = p_1 $ if $k = 1$ and 
\begin{align*}
 P(Y_{i,k}=1) &= P(\tilde{Y}_{i,k} =1) -  P(\tilde{Y}_{i,k-1} =1) \\
  &=  p_1 + \sum_{q=2}^{k} p_q \prod_{m=1}^{q-1} (1-p_m) -  \left(p_1 + \sum_{q=2}^{k-1} p_q \prod_{m=1}^{q-1} (1-p_m)\right) \\
  &= p_{k} \prod_{m=1}^{k-1} (1-p_m),
\end{align*}
otherwise. On the other hand, the probability that a patient does not show toxicity at a given cycle  $k$ is given by the probability to not have shown any toxicity until  $k$, that is,
\begin{align*}
 P(Y_{i,k}=0)  &=  \prod_{m=1}^{k} (1 - p_m)  \\
  &= 1 -   P(\tilde{Y}_{i,k}=1).
\end{align*}

\subsection*{A.2.\enspace Scenario details}\label{app2}
In the following, Table~\ref{tablescen} shows the generated 
cumulative probabilities of toxicity for each scenario, dose sequence and cycle.

\begin{table}
\begin{center}
\caption{Cumulative probabilities of toxicity for each scenario.}\label{tablescen}
\begin{tabular}{lcccccccccccl}
\hline
\textbf{Dose}   & \multicolumn{5}{c}{\textbf{Cycle}} &  & \multicolumn{5}{c}{\textbf{Cycle}} \\ 
\textbf{sequence}  & 1& 2 & 3 & 4 & 5  &  & 1& 2 & 3 & 4 & 5  \\
\hline
 &\multicolumn{5}{c}{Scenario 1} & & \multicolumn{5}{c}{Scenario 2}\\
$\bar{s}_1$ & 0.05 & 0.06 & 0.07 & 0.08 & 0.11 & & 0.01 & 0.01 & 0.01 & 0.02 & 0.03 \\
$\bar{s}_2$ &0.07 & 0.08 & 0.10 & 0.15 & 0.20 & &0.01 & 0.02 & 0.02 & 0.03 & 0.06 \\
$\bar{s}_3$ &0.10 & 0.12 & 0.15 & 0.20 & 0.30 & &0.02 & 0.02 & 0.04 & 0.10 & 0.15 \\
$\bar{s}_4$ &0.14 & 0.18 & 0.25 & 0.36 & 0.50 & &0.03 & 0.04 & 0.07 & 0.14 & 0.30  \\
$\bar{s}_5$ &0.18 & 0.25 & 0.35 & 0.50 & 0.65 & &0.04 & 0.06 & 0.11 & 0.25 & 0.50 \\
\\
 &\multicolumn{5}{c}{Scenario 3} & & \multicolumn{5}{c}{Scenario 4}\\
$\bar{s}_1$ & 0.04 & 0.05 & 0.05 & 0.05 & 0.06 & & 0.15 & 0.15 & 0.16 & 0.18 & 0.20 \\
$\bar{s}_2$ &0.06 & 0.06 & 0.07 & 0.08 & 0.09 & &0.23 & 0.23 & 0.25 & 0.28 & 0.30 \\
$\bar{s}_3$ &0.08 & 0.09 & 0.10 & 0.12 & 0.15 & &0.33 & 0.35 & 0.38 & 0.40 & 0.45\\
$\bar{s}_4$ &0.12 & 0.13 & 0.16 & 0.18 & 0.22 &  &0.48 & 0.50 & 0.54 & 0.60 & 0.65  \\
$\bar{s}_5$ &0.15 & 0.17 & 0.20 & 0.24 & 0.30 & &0.58 & 0.62 & 0.66 & 0.72 & 0.80 \\
\\
 &\multicolumn{5}{c}{Scenario 5} & & \multicolumn{5}{c}{Scenario 6}\\
$\bar{s}_1$ &0.14 & 0.15 & 0.18 & 0.23 & 0.30 & & 0.10 & 0.10 &0.10 &0.10 &0.10 \\
$\bar{s}_2$ &0.21 & 0.24 & 0.29 & 0.38 & 0.50 & &0.20 & 0.20 &0.20 &0.20 &0.20 \\
$\bar{s}_3$ &0.31 & 0.36 & 0.45 & 0.58 & 0.70 & &0.30 & 0.30 &0.30 &0.30 &0.30 \\
$\bar{s}_4$ &0.45 & 0.53 & 0.66 & 0.79 & 0.85 & &0.40 & 0.40 &0.40 &0.40 &0.40 \\
$\bar{s}_5$ &0.56 & 0.66 & 0.78 & 0.89 & 0.95 & &0.50 & 0.50 &0.50 &0.50 &0.50 \\
\hline
\end{tabular}
\end{center}
\end{table}

\subsection*{A.3.\enspace Fernandes et al.'s results}\label{app3}
Let $p_{i,k}=P(Y_{(i,k)}=1 | Y_{(i,k-1)}=0,\ldots, Y_{(i,1)}=0$) be the conditional probability of toxicity given the absence of any toxicity at the previous cycles. \cite{fernandes2016b} proposed a conditional modeling approach to phase I adaptive clinical trials where the probability of toxicity at each cycle is modeled via a log-linear link, that is,

\begin{equation}
p_{i,k}=1-\exp[-\alpha(d_{i,k}-\rho \dot{d}_{i,k})^+ -\beta \tilde{D}_{i,k}d_{i,k}],
\end{equation}
where $d_{i,k}$ and $\tilde{D}_{i,k}= \sum_{m=1}^{k-1}d_{i,m}$ are the actual dose and the cumulative dose that were administered to patient $i$ at the cycle $k$, respectively; $\dot{d}_{i,k}$ denotes the maximum dose already given to the $i$th patient in previous cycles and $(x)^+$ is equal to $x$ if $x>0$ and 0 otherwise.
$\alpha$ is the effect of the administered dose at the first cycle, $\beta$ that of the cumulative dose, and $0 \leq \rho \leq 1$ reflects the amount of memory regarding whether a previous maximum dose was tolerable. 

Following the details described in Section 3.1 of \cite{fernandes2016b}, we run 1000 simulations in two scenarios. The first is the one suggested by the authors, with 30 patients equally distributed within 5 doses (with the skeleton of (0.02, 0.05, 0.10, 0.15, 0.23)), 6 cycles and $\alpha=1$, $\beta = 0.5$ and $\rho = 0.8$. We run the model at the end of each simulated trial, that is, when all patients have ended the follow-up period. The JAGS code suggested in the manuscript was used along with the prior distributions: $\alpha \sim \mathcal{LN}(-0.8047190,0.6213349)$, $\beta\sim \mathcal{LN}(-1.498,0.621)$ and $\rho \sim Beta(5, 1)$, with $\mathcal{LN}$ denoting the lognormal distribution parameterized by the precision parameter.
In a second scenario, we set $\rho = 0.4$.

Table~\ref{tablefern} shows the results in terms of parameter estimation and bias. $\rho$ is the more difficult parameter to be estimated, and the prior distribution plays a crucial role.

\begin{table}
\begin{center}
\caption{Results in terms of parameter estimation and bias in two scenarios for the model proposed by \cite{fernandes2016b}.}\label{tablefern}
\begin{tabular}{lccc}
\hline
\textbf{Parameter}   & \textbf{True value} & \textbf{Estimation}  & \textbf{Bias}  \\ 
 & & median (Q1, Q3)  & median (Q1, Q3) \\
\hline
Proposed scenario & & & \\
$\alpha$ & 1 & 0.939(0.681,1.27) & -0.061(-0.319,0.27) \\ 
$\beta$ & 0.5 & 0.437(0.32,0.647) & -0.063(-0.18,0.147) \\ 
$\rho$  &  0.8 & 0.809(0.773,0.843) & 0.009(-0.027,0.043) \\ 
& & & \\
Modified scenario & & & \\
$\alpha$  &  1 &1.151(0.845,1.532) & 0.151(-0.155,0.532) \\ 
$\beta$  &  0.5 & 0.733(0.486,1.116) & 0.233(-0.014,0.616) \\ 
$\rho$  &  0.4 & 0.727(0.668,0.774) & 0.327(0.268,0.374) \\ 
\hline
\end{tabular}
\end{center}
\end{table}


\begin{thebibliography}{}

\bibitem[Braun et~al., 2007]{braun07}
Braun, T.~M., Thall, P.~F., Nguyen, H., and De~Lima, M. (2007).
\newblock Simultaneously optimizing dose and schedule of a new cytotoxic agent.
\newblock {\em {Clinical Trials}}, 4(2):113--124.

\bibitem[Braun et~al., 2005]{braun05}
Braun, T.~M., Yuan, Z., and Thall, P.~F. (2005).
\newblock Determining a maximum-tolerated schedule of a cytotoxic agent.
\newblock {\em Biometrics}, 61(2):335--343.

\bibitem[Cheung and Chappell, 2000]{cheung2000}
Cheung, Y. and Chappell, R. (2000).
\newblock Sequential designs for phase {I} clinical trials with late-onset
  toxicities.
\newblock {\em Biometrics}, 56(4):1177--1182.

\bibitem[Cheung, 2011]{Cheung_book}
Cheung, Y.~K. (2011).
\newblock {\em Dose Finding by the Continual Reassessment Method}.
\newblock Chapman \& Hall/CRC Biostatistics Series.

\bibitem[Fernandes et~al., 2016]{fernandes2016b}
Fernandes, L.~L., Taylor, J.~M., and Murray, S. (2016).
\newblock {Adaptive phase {I} clinical trial design using Markov models for
  conditional probability of toxicity}.
\newblock {\em Journal of Biopharmaceutical Statistics}, 26(3):475--498.

\bibitem[Iasonos and O'Quigley, 2014]{iasonos2014}
Iasonos, A. and O'Quigley, J. (2014).
\newblock Adaptive dose-finding studies: A review of model-guided phase {I}
  clinical trials.
\newblock {\em Journal of Clinical Oncology}, 32:2505--2511.

\bibitem[Lee and Cheung, 2009]{lee2009}
Lee, S.~M. and Cheung, Y.~K. (2009).
\newblock Model calibration in the continual reassessment method.
\newblock {\em Clinical Trials}, 6(3):227--238.

\bibitem[Legedza and Ibrahim, 2010]{legedza2000}
Legedza, A. and Ibrahim, J. (2010).
\newblock Longitudinal design for phase {I} clinical trials using the continual
  reassessment method.
\newblock {\em Controlled Clinical Trials}, 21:574--588.

\bibitem[Liu and Braun, 2009]{liu09}
Liu, C.~A. and Braun, T.~M. (2009).
\newblock Parametric non-mixture cure models for schedule finding of
  therapeutic agents.
\newblock {\em {Journal of the Royal Statistical Society: Series C (Applied
  Statistics)}}, 58(2):225--236.

\bibitem[Lyu et~al., 2018]{lyu18}
Lyu, J., Curran, E., and Ji, Y. (2018).
\newblock Bayesian adaptive design for finding the maximum tolerated sequence
  of doses in multicycle dose-finding clinical trials.
\newblock {\em JCO Precision Oncology}, 2:1--19.

\bibitem[Morita et~al., 2008]{Morita08}
Morita, S., Thall, P.~F., and Muller, P. (2008).
\newblock {{D}etermining the effective sample size of a parametric prior}.
\newblock {\em Biometrics}, 64(2):595--602.

\bibitem[Neuenschwander et~al., 2008]{Neuenschwander08}
Neuenschwander, B., Branson, M., and Gsponer, T. (2008).
\newblock {Critical aspects of the Bayesian approach to phase {I} cancer
  trials}.
\newblock {\em Statistics in Medicine}, 27(13):2420--2439.

\bibitem[O'Quigley et~al., 1990]{o1990}
O'Quigley, J., Pepe, M., and Fisher, L. (1990).
\newblock Continual reassessment method: a practical design for phase 1
  clinical trials in cancer.
\newblock {\em Biometrics}, pages 33--48.

\bibitem[Paoletti et~al., 2015]{paoletti2015}
Paoletti, X., Doussau, A., Ezzalfani, M., Rizzo, E., and Thi\'{e}baut, R.
  (2015).
\newblock Dose finding with longitudinal data: simpler models, richer outcomes.
\newblock {\em Statistics in Medicine}, 34(22):2983--2998.

\bibitem[Postel-Vinay et~al., 2014]{postel14}
Postel-Vinay, S., Collette, L., Paoletti, X., Rizzo, E., Massard, C., Olmos,
  D., Fowst, C., Levy, B., Mancini, P., Lacombe, D., et~al. (2014).
\newblock {Towards new methods for the determination of dose limiting
  toxicities and the assessment of the recommended dose for further studies of
  molecularly targeted agents -- Dose-Limiting Toxicity and Toxicity Assessment
  Recommendation Group for Early Trials of Targeted therapies, an European
  Organisation for Research and Treatment of Cancer-led study}.
\newblock {\em European Journal of Cancer}, 50(12):2040--2049.

\bibitem[Wages et~al., 2014]{wages14}
Wages, N.~A., O'Quigley, J., and Conaway, M.~R. (2014).
\newblock Phase {I} design for completely or partially ordered treatment
  schedules.
\newblock {\em Statistics in Medicine}, 33(4):569--579.

\bibitem[Zhang and Braun, 2013]{zhang13}
Zhang, J. and Braun, T.~M. (2013).
\newblock A phase {I} {B}ayesian adaptive design to simultaneously optimize
  dose and schedule assignments both between and within patients.
\newblock {\em Journal of the American Statistical Association},
  108(503):892--901.

\bibitem[Zohar and O'Quigley, 2006]{zohar2006}
Zohar, S. and O'Quigley, J. (2006).
\newblock Optimal designs for estimating the most successful dose.
\newblock {\em Statistics in Medicine}, 25:4311--4320.

\end{thebibliography}
\end{document}